\documentclass{aa}
\usepackage{txfonts}
\usepackage{graphicx}
\usepackage{natbib}
\bibpunct{(}{)}{;}{a}{}{,}

\usepackage{xcolor}

\usepackage[normalem]{ulem}

\begin{document}

\title{Misaligned tidal tails in open clusters: a signature of the local 
spiral arm resonance}

\author{Andr\'es E. Piatti\inst{1,2,}\thanks{\email{andres.piatti@fcen.uncu.edu.ar}}}

\institute{Instituto Interdisciplinario de Ciencias B\'asicas (ICB), CONICET-UNCuyo, Padre J. Contreras 1300, M5502JMA, Mendoza, Argentina;
\and Consejo Nacional de Investigaciones Cient\'{\i}ficas y T\'ecnicas (CONICET), Godoy Cruz 2290, C1425FQB,  Buenos Aires, Argentina\\
}

\date{Received / Accepted}

\abstract
{Tidal tails of open clusters are traditionally expected to align with their orbital motion 
due to Galactic shear. However, recent observations suggest that a significant population of 
clusters exhibits misaligned tails, the origin of which remains poorly understood.}
{I investigate the structural properties of tidal tails for a sample of 155 nearby open 
clusters within 1 kpc of the Sun to characterize the influence of non-axisymmetric Galactic 
perturbers on stellar stripping geometry.}
{Using high-precision {\it Gaia} DR3 astrometry and the probabilistic membership, I trace 
the primary orientation and physical length of tidal tails. I distinguish between local and 
global perturbations by analyzing the spatial distribution of misalignment ($\Delta$PA) 
relative to the boundaries of the Local Arm and comparing results with recent test-particle 
simulations of barred and spiral potentials.}
 {I identify three distinct dynamical regimes. While the majority of the population is 
shear-aligned, a significant subset exhibits a spatially indifferent background misalignment 
(10$^o \le \Delta$PA $\le$ 30$^o$) consistent with global Galactic bar torques.  In contrast, 
a population of nascent tails ($\le$ 0.3 kpc) exhibits larger misalignments ($\Delta$PA 
$\gtrsim$ 40$^o$) with a unique spatial specificity, occurring exclusively within the 
inter-arm gaps ($|d_{arm}| \gtrsim$ 0.2 kpc).}
  {These significant misalignments could represent a transient dynamical phase occurring within 
the corotation resonance's zero-shear zone, where the radial gravitational forcing of the 
spiral potential dominates stellar exit trajectories by rotating the eigenvectors of the 
tidal tensor. This could identify nascent tidal tails as uniquely sensitive high-resolution 
tracers of the non-axisymmetric Galactic potential.}

\keywords{Methods: data analysis -- (Galaxy:) open clusters and associations: general}

\titlerunning{Open cluster tidal tails}

\authorrunning{A.E. Piatti}           

\maketitle

\markboth{A.E. Piatti:}{Open cluster tidal tails}

\section{Introduction}       

Tidal tails in open clusters have increasingly been detected  since recent time
\citep[see, e.g.,][]{bhattacharyaetal2022}. Observed properties of tidal tails, 
such as morphology, binary fraction, dynamics, rotation, etc \citep{weisetal2025,sharmaetal2025},
as well as simulated tidal tails \citep{pflamm-altenburgetal2023,kroupaetal2024,jadhavetal2025} 
have fueled this field of research. Among these efforts, a catalogue of tidal tails for 476 
open clusters distributed within a circle of 3 kpc from the Sun has been built by 
\citet[][see also references therein]{kos2024}. As far as I am aware, this is the largest 
compilation of open cluster tidal tails. The catalogue includes mean open cluster astrophysical 
parameters, namely: Galactrocentric coordinates, radial velocity, proper motion, age, etc, and 
the corresponding ones for the cluster and tidal tail members with their respective 
membership probabilities. The catalogue has become a wealth of opportunities
in order to exploit it with the aim of performing for the first time a meaningful 
statistical analysis of the open cluster tidal tails, and, more importantly, to search for 
localized dynamical anomalies that challenge our current understanding of stellar stripping.
 
The spatial orientation of these tidal tails provides a direct window into the interaction 
between open clusters and the Milky Way potential. Theoretically, tidal tails are expected to 
align primarily along the cluster’s orbit, with a leading arm pulling toward the Milky Way 
center and a trailing arm extending outward \citep{montuorietal2007}. However, this study 
reports the discovery of a distinct population of clusters where this expectation fails. I 
isolate a unique localized dynamical signature that reveals how the gravitational field of 
the Local Arm influences the geometry of stellar stripping near the corotation radius.
In this zero-shear zone, the suppression of standard azimuthal stretching is hypothesized 
to allow radial gravitational perturbations to dominate the stripping process 
\citep{barrosetal2021}.

Furthermore, the physical length of tidal tails serves as a chronological record of a 
cluster’s mass-loss history. During the cluster's lifetime, continuous internal two-body 
relaxation and external tidal shocks drive stars beyond the tidal radius, gradually 
extending the structures over hundreds of millions of years \citep{kupperetal2010}. 
I propose that shorter tails ($<$0.5 kpc) act as instantaneous tracers of the current 
local gravitational gradient. This perspective could allow us to distinguish between 
the time-averaged alignment of older clusters and the transient, non-aligned response 
of younger systems.
While younger clusters (typically $<$ 100 Myr)  may exhibit only nascent, high-density protrusions, 
more evolved systems are expected 
to possess extensive, diffuse tails that can span several hundred parsecs. A statistical 
comparison between cluster age and tail length  allows us to test whether 
the observed expansion of these tails scales consistently with cluster evolution or is heavily 
dependent on the specific environment of the solar neighborhood.

Building upon the groundwork laid by recent surveys and the comprehensive data provided by 
\citet{kos2024}, the present work aims to characterize a physically distinct population of 
misaligned clusters. Specifically, I investigate the correlations between tail length,
cluster age,  and global orientation in the Milky Way disc to determine if the physical 
extension of these  structures dictates their susceptibility to the Local Arm resonance. 
By performing this 
statistical analysis, I seek to provide a benchmark for current N-body simulations and improve 
our understanding of the role open clusters play in the secular building of the Milky Way's 
thin disc. Section~2 deals with handling and describes the data analysis, which includes
computed tidal tail lengths and position angles. Section~3  presents the discovery of 
the non-aligned population and discusses its connection to the Local Arm resonance.
I summarize in Section~4 the main conclusions of this work.

\section{Data handling}

I based the statistical analysis on the properties of 1.045.225 stars included
in the \citet{kos2024}'s catalogue. The catalogue was built from the {\it Gaia} DR3
data base \citep{gaiaetal2016,gaiaetal2022b},  using stars brighter than
$G$ = 17.5 mag. I refer the reader to \citet{kos2024} for details concerning 
membership probabilities of stars to the open clusters or to the tidal tails, and
the astrophysical parameters derived for them. Specifically, I used the Galactocentric
coordinates ($X,Y,Z$), and the recommended membership probabilities of the stars to belong 
to the main body of the clusters or to their tidal tails. Since the errors in the
heliocentric distances are proportional to the square of the heliocentric distance
($\sigma$$_d$ $\propto$ $d^2$$\sigma$$_\varpi$), I limited
the sample to open clusters located within a circle of 1 kpc from the Sun, so that
the errors in the Galactocentric coordinates for the farthest open clusters resulted
to be smaller than $\sim$ 50 pc \citep{lindegrenetal2021a,lindegrenetal2021b}, thus
ensuring that any detected misalignments are physically real and not artifacts of 
distance-dependent errors.
The total number of open clusters at $d$ $\le$ 1 kpc from the Sun in the \citet{kos2024}'s
catalogue downs to 155, providing the high-precision framework necessary to distinguish 
between standard orbital alignment and the anomalous paths reported here. The Milky Way 
framework is one with the $X$ axis increasing in the direction from the Sun to the Milky Way
centre, and the $Y$ axis perpendicular to it and increasing toward the left of
the Milky Way centre as seen from the Sun. In this framework, the Milky
Way rotation is clockwise, seen from the North Galactic Pole.

\begin{figure}
\includegraphics[width=\columnwidth]{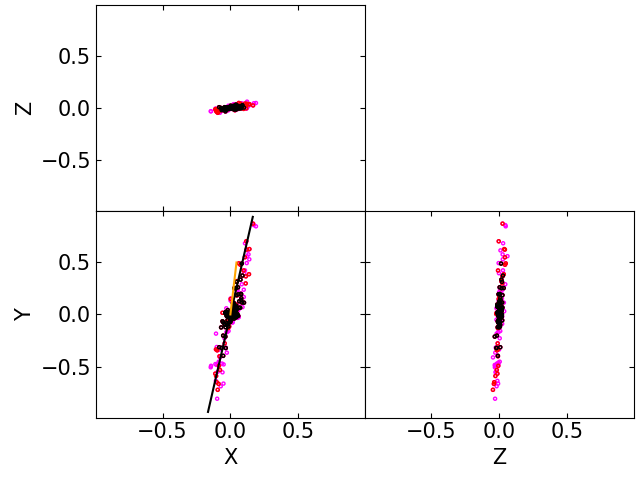}
\caption{Galactic coordinate planes (in kpc) for stars in the field of the open cluster
NGC~1342. The coordinate framework has its origin in the open cluster's centre. Black, 
red, and magenta symbols represent stars with $P$ $>$ 90\%,
70\% $<$ $P$ $<$ 90\%, and 50\% $<$ $P$ $<$ 70\%, respectively. The black solid
line represents the traced tidal tail, while the orange one represents the cluster's 
direction of motion.}
\label{fig1}
\end{figure}

The robustness of this 155 cluster sample is underscored by the high completeness of the 
{\it Gaia} DR3 census within the first kiloparsec. Recent work has established the first 
formal selection function for this census, demonstrating that cluster detectability can 
be modeled with 94.5\% accuracy based on parameters such as the number of stars, median 
parallax error, and local Gaia data density \citep{huntetal2026}. Note that the 
94.5\% accuracy refers specifically to the completeness model of the selection function 
from \citet{huntetal2026}, rather than representing a false positive rate of cluster 
identification. This selection model 
reveals that cluster detectability depends on factors such as proper motion, where even 
modest boosts in orbital speed can increase detection probability. {}
Complementing this, recent examinations of Galactic open clusters using {\it Gaia} DR3 
have identified approximately 40\% more stars in systems within 1 kpc compared to previous 
DR2 based catalogs \citep{alfonsoetal2024}. This improvement is largely due to the systematic inclusion of faint members ($G > 17$ mag) and the application of more accurate parallax zero-point corrections, which reduce artificial stretching in the spatial distribution and allow for more reliable membership identification. Within this specific volume, 
the average parallax error is reduced to approximately 0.16 mas, providing a precise 
astrometric foundation for tracing the geometry of extended tidal structures. By utilizing 
a sample drawn from this highly complete and expanded census, I ensure that the search 
for localized dynamical anomalies is conducted within a statistically sound framework. 
This high precision is essential for distinguishing between the standard, shear-aligned 
population, which could serve as a robust control group, and the unique localized dynamical 
signature of misaligned tails that constitutes our primary discovery.

Membership probabilities ($P$) have been thoroughly estimated by \citet{kos2024},
who found that stars with $P$ $>$ 90$\%$  satisfactorily match those of open clusters 
with previously detected tidal tails. As expected, these stars do not make up tidal tails
as extended as those composed by stars with $P$ $<$ 50$\%$, as seen in
\citet{kos2024}. Statistically speaking, stars with $P$ = 50$\%$ have the same chance 
to belonging to the cluster's tidal tails or to the Milky Way field. This means that 
considering stars with $P$ $<$ 50$\%$ undoubtedly could lead to misleading results. 
In contrast to \citet{kos2024}, who  traced tidal tails on the basis of stars with any membership
probability, I preferred to trace physically meaningful tidal tails by employing only stars with 
$P$ $>$ 50$\%$. In practice, I considered three $P$ intervals, namely: $P$ $>$ 90\%,
70\% $<$ $P$ $<$ 90\%, and 50\% $<$ $P$ $<$ 70\%, respectively, so that I monitored
the shape, orientation and length of the tidal tails in terms of the $P$ values. 

\begin{figure}
\includegraphics[width=\columnwidth]{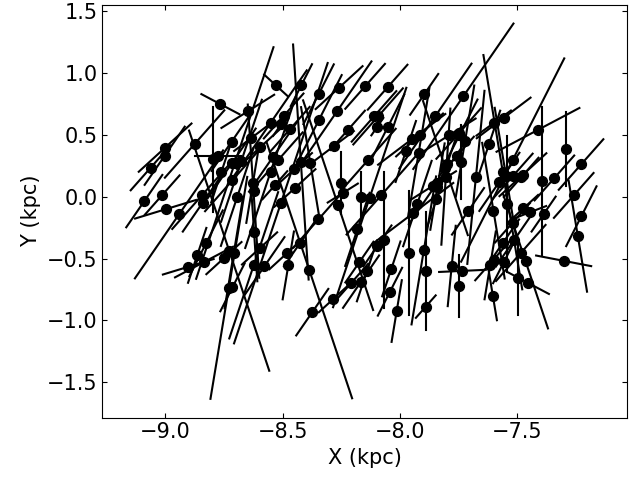}
\caption{Positions of the studied open clusters in the Milky Way plane (black filled circles),
and their corresponding measured tidal tails (black solid lines).}
\label{fig2}
\end{figure}

This probabilistic approach represents a significant methodological departure from 
kinematics-based techniques, such as the compact convergent point (CCP) method developed 
by \citet{jerabkovaetal2021}. The classical CCP method filters field contamination by 
decomposing stellar proper motions into parallel ($v_\parallel$) and perpendicular 
($v_\perp$) components relative to the cluster's 3D bulk velocity vector 
$\mathbf{v}_{\text{cls}}$. To account for the spatial extension of the tails under 
standard Galactic shear, the CCP method applies a linear distance-dependent correction:\\

$    v_{\parallel, \text{corr}} = v_{\parallel, \text{sim}} - v_{\parallel, \text{pred}} - \Gamma(R - R_{\text{cl}}),$\\

\noindent where $R - R_{\text{cl}}$ is the heliocentric distance separation from the cluster 
centre and $\Gamma = \partial(\Delta v_\parallel)/\partial R$ represents the linear velocity 
gradient along the stream. Stars are classified as members if they fall within a tight 
kinematic envelope: $|v_{\parallel, \text{corr}}| < \sigma_\parallel$ and $|v_\perp| < \sigma_\perp$.

While highly effective for shear-dominated streams, the rigid linear formulation of the 
CCP method could become systematically biased when applied to clusters residing within or 
near resonance regions. In the zero-shear zone of the Local Arm's corotation resonance, 
standard azimuthal stretching is minimized, and the linear shear gradient breaks 
down ($\Gamma \to 0$). Instead of steady-state drifting, escaping stars perform
 non-linear horseshoe orbits or coherent breathing motions 
\citep{barrosetal2021, asanoetal2024}. Consequently, their velocities deviate 
strongly from the expected linear shear relation, causing standard kinematic cuts to 
discard true, highly misaligned members as field contaminants. 

In contrast, the probabilistic framework of \citet{kos2024} evaluates membership by 
comparing a forward-modeled N-body likelihood $L(\mathbf{w})$ across 6D phase-space
 $\mathbf{w} = (x, y, z, v_x, v_y, v_z)$ against an empirical background field density 
$F(\mathbf{w})$:\\

$    P(\text{member} \mid \mathbf{w}_i) = \frac{\eta L(\mathbf{w}_i)}{\eta L(\mathbf{w}_i) + (1 - \eta) F(\mathbf{w}_i)},$\\

\noindent where $\eta$ is a localized prior. Because this Bayesian formulation does not enforce a 
rigid velocity-distance relation ($\Gamma$), it naturally accommodates the highly non-linear,
 rotated exit trajectories of nascent tails ($\le 0.3$~kpc) in the inter-arm gaps.
 Grounding our sample on these unconstrained membership probabilities ensures a 
contamination-minimized physical representation of the tails, preserving the crucial 
non-axisymmetric signatures ($\Delta\mathrm{PA} \ge 40^\circ$) that would otherwise be
 filtered out by standard linear kinematic algorithms.

\begin{figure}
\includegraphics[width=\columnwidth]{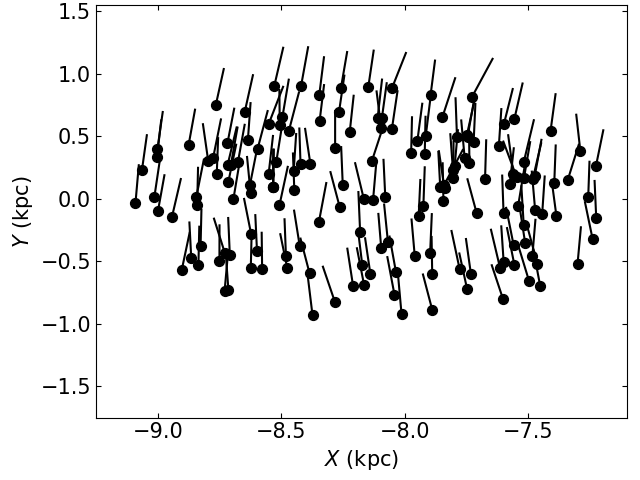}
\caption{Positions of the studied open clusters in the Milky Way plane (black filled circles),
and their corresponding velocity direction (black solid lines).}
\label{fig3}
\end{figure}

Figure~\ref{fig1} shows an example of the recovered open clusters' tidal tails. 
The coordinate framework has its origin in the open cluster's centre \citep{kos2024}. 
Figures similar to Figure~\ref{fig1} for the complete sample of open clusters' tidal 
tails are provided as online Supplementary material.
In Figure~\ref{fig1}, stars with $P$ $>$ 90\%, 70\% $<$ $P$ $<$ 90\%, and 50\% $<$ $P$ $<$ 70\%
are drawn with black, red, and magenta open circles, respectively. As can be seen,
while black symbols strongly delineate the tidal tails emerging from the cluster's body,
the red and magenta ones trace them even further. For some open clusters, the innermost 
regions of the tidal tails seem to have an orientation that differs from that of the
outermost ones. This is the typical S-shaped path of stars evaporating from star clusters
\citep[see, e.g.,][]{pflamm-altenburgetal2023,pflamm-altenburg2025}, which is
due the epicyclic motion of stars after evaporation from the star clusters with a superimposed
secular evolution in y-direction \citep[e.g.,][]{kupperetal2008}, leading to a regular pattern 
of over-densities \citep[e.g.,][Eq. (8)]{kupperetal2010}.

\subsection{Methodology for tracing tidal tail geometry}

While tidal tails are inherently curved due to epicyclic motion (see, Figure~\ref{fig1}), 
their global 
orientation can be effectively linearized to determine their primary alignment relative 
to the Galactic potential. This approach is particularly robust for identifying the 
instantaneous trajectories of shorter tails ($< 0.5$ kpc), ensuring that the 
tidal tails orientation reflects the immediate dynamical response to the local potential before 
orbital curvature or shear-driven stretching occurs.
I determined the geometry of these structures using the following objective criteria:
1) The coordinate system for each cluster was first centred on the cluster’s 
core ($X_c$,$Y_c$) as defined in the \citet{kos2024}'s catalogue. This origin serves as the vertex 
for the leading and trailing arm vectors. 2) I did not treat all stars equally. I 
firstly used stars 
with $P$ $>$ 90\% (black symbols in Figure~\ref{fig1}) to define the initial exit direction of the 
tails from the Lagrange points. Stars with 50\% $<$ $P$ $<$ 90\% (red and magenta symbols) were 
then used to trace the secular evolution of the tails into the field.
In other words,  $P>$90\% stars act as an anchor to define the initial exit trajectory at the 
Lagrange points, while the linear regression (see below) is performed on the binned medians 
of the entire $P>$50\% sample. 3) To obtain the position angle (PA), 
I performed a linear fit to the ($X,Y$) coordinates of all members with $P$ $>$ 50\%. However, 
to avoid the fit being skewed by local over-densities near the cluster core, I applied a spine-tracing
technique: I binned the stars along the perceived tail and calculated the median ($X,Y$) position 
within each bin. The bin width was set dynamically based on the cluster's tidal tail extension, but typically averaged a physical width of 0.05 kpc to ensure a sufficient number of member 
stars per bin to compute a robust median ($X,Y$) position.
The final straight line is the best fit through these median points. 4)
The PA is the angle measured from the positive $X$ axis (pointing toward the Galactic centre) 
towards the straight line that best represents the total extension of the tidal tails.
A PA of 90$\degr$ would indicate a tail perfectly aligned with the $Y$ axis. 5) The total length
was calculated as the Euclidean distance between the two most distant stars ($P$ $>$ 50\%) found along 
the leading and trailing spine. In Figure 1, for NGC~1342, the line represents this maximal extent, 
reaching approximately 0.45 kpc. Uncertainties were estimated by slightly varying the membership probability 
threshold (e.g., $P$ $>$ 60\%) and recalculating the PA and length. I found that the orientations and
lenghts are robust, with typical variations of $\pm$5$\degr$ and $\pm$0.05 kpc, respectively.
Table~\ref{tab1} lists the adopted mean PA and length values.
By adopting this method, I capture the projected primary orientation of the stellar spine, which serves as the most relevant parameter for distinguishing between instantaneous local perturbations 
and the time-averaged effect of global Galactic shear.

\begin{figure}
\includegraphics[width=\columnwidth]{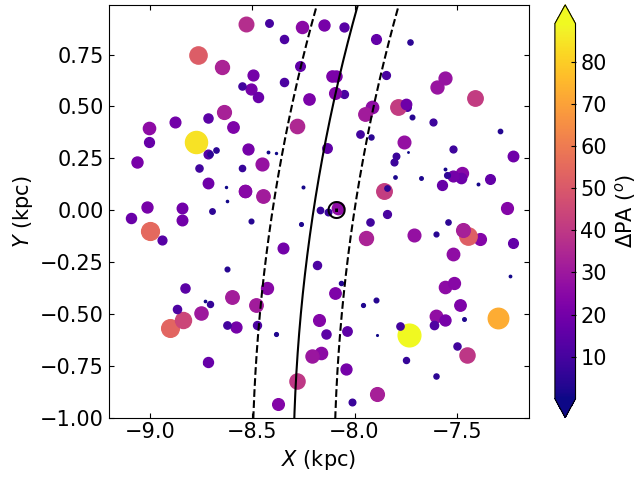}
\caption{Positions of the studied open clusters in the Milky Way plane coloured
according to the $\Delta$PA values. The size of the symbols are also 
proportional to $\Delta$PA. The solid and dashed lines represent the mean 
position and width of the Local Arm \citep{haoetal2021}. The Sun is placed at 
($X$= -8.12,$Y$=0.0).}
\label{fig4}
\end{figure}

\section{Analysis and discussion}

I built Figure~\ref{fig2} with the aim of visualizing the spatial distribution
of the tidal tail lengths and position angles in the Milky Way plane. As can be
seen, the spatial distribution pattern of the position angles  is not the
expected one, where the shear of the disc makes closer or further stars
to the Milky Way centre to move faster or slower, respectively,  
Because of the inner part of an open cluster moves faster 
than the outer part, tidal tails are naturally sheared into tangential arcs.
This shear is what stretches tidal tails into tangential arcs along the
open cluster orbits, which mostly follow the rotation of the Milky
Way \citep{tarricgetal2021}. For the sake of the reader, I show in
Figure~\ref{fig3} the directions of motion of the open clusters. They
were computed using the mean open clusters' proper motions, radial velocities,
and heliocentric distances in \citet{kos2024}. Following the recipe for deriving 
the PAs of the tidal tails, I also derived the PAs of the
open clusters' velocity vectors.
From this point of view, I defined  the  PA difference 
or misalignment between these two vectors ($\Delta$PA = $|$PA velocity vector 
-  PA tidal tails$|$). Figure~\ref{fig4} shows the positions of the open clusters
in the Milky Way plane. From the figure, I report the discovery of a distinct 
population of clusters whose tidal tails are misaligned with their 
directions of motion $\Delta$PA $\gtrsim$ 20$^o$), representing a departure from standard 
shear-driven expectations. Note that because the spine-fitting of tidal tails
(see Section 2.1) is anchored in the inner high-density regions and is highly 
stable under varying membership thresholds, the incomplete detection of faint outer 
members does not significantly affect the calculated global misalignment angles.

\begin{figure}
\includegraphics[width=\columnwidth]{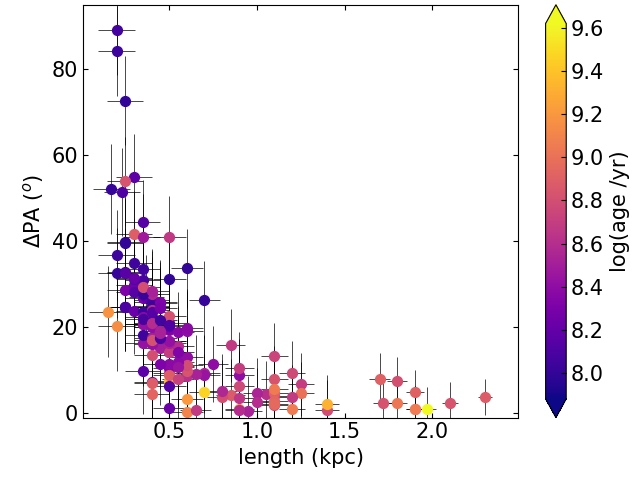}
\caption{Relationship between $\Delta$PA and the length of the tidal
tails, with their error bars, coloured according to the respective clusters' ages.}
\label{fig5}
\end{figure}

In recent years, the corotation radius associated with the Local Arm, and hence the confirmation 
of the position of the Sun near to it, has played a increasing relevant role in studies of the structure and kinematics of the Milky Way disc \citep{diasetal2019,liuetal2025,bobylevetal2025,kg2025}. 
I note that while the Milky Way hosts multiple spiral arms with potentially different dynamics and 
pattern speeds \citep{roskaretal2012,reidetal2019}, the present analysis focuses on the 
zero-shear zone created by the resonance most relevant to the cluster sample (those near the Local Arm). The aforementioned corotation radius is then the specific distance from the
Milky Way centre where stars and gas orbit at 
the exact same speed (angular velocity, $\Omega$) as the Local Arm \citep{barrosetal2021};
the latter moving according  to the spiral density wave theory \citep{pour-imanietal2016}.

I investigated whether the particular location of the studied open clusters in the Milky Way
could be a possible explanation for the tidal tail orientation pattern shown in Figures~\ref{fig2}
and \ref{fig4}. By placing these clusters where azimuthal shear is suppressed, I show that 
radial perturbations from the spiral potential would seem to become the primary driver of stellar exit 
from the Lagrange points.
At the corotation radius, the stretching force that usually makes tidal tails tangential 
would be minimized. If an open cluster is trapped in the potential well at corotation, 
the escaped stars would not drift away along the orbit as quickly. 
In this zero shear zone, 
without the stretching force to pull the tidal tails into tangential arcs along the orbit, the 
primary remaining force would be the radial  perturbation from the spiral arms themselves, so that stars 
escaping an open cluster would be more likely to be pushed or pulled radially by the spiral potential 
rather than being smeared into a long tangential ribbon. 

At the corotation radius, stars do not follow simple circles. Instead, 
they perform horseshoe orbits in the rotating frame of the spiral pattern. The stars oscillate 
radially inward and outward as they interact with the spiral arm’s gravitational potential. 
\citet{fujiibaba2012} and \citet{asanoetal2024} have identified disc stars with coherent, radial, 
and vertical oscillations, known as breathing motions \citep{gaiaetal2023,liuetal2025}, 
in the Local Arm among others. In this context, the misaligned tidal tails I observe may be 
tracing these complex local flows rather than simple circular orbits.

Figure~\ref{fig5} depicts the relationship between $\Delta$PA with the length of the
tidal tails and the respective clusters' ages. According to \citet{haoetal2021}, the Local 
Arm where the Sun is 
currently located, as well as the other nearly spiral arms, could be long-lived arms of the 
Milky Way. Older clusters have likely
undergone multiple disc crossing and significant migration \citep{jerabkovaetal2021},
while younger ones has had  less time to deviate from their birth environment,
so that they have remained in closer proximity to their natal spiral structures.
Hence, I explore the possibility that the observed orientations of cluster tidal 
tails represent a transient response to this sustained local potential before these systems 
eventually migrate or conform to the global Galactic shear.

Recently, \citet[][and references therein]{dinnbieretal2022}
argued that the extension of the tidal tails is linked to the parent cluster's age, in the sense
that the older an open cluster, the longer its tidal tails. Figure~\ref{fig5} reveals 
 a general trend of the length of tidal tails with age, and a spread of
ages for relative short tidal tails ($<$ 0.5 kpc). Moreover, relative short tidal tails
seem to correlate with $\Delta$PA, in the sense that the shorter the tidal tails the
larger the $\Delta$PA values.  Considering the inherent uncertainties, the increasing 
misalignment seen in Figure~\ref{fig5}
for shorter tidal tails arises as an apparent feature of the studied cluster sample.
I note that regardless of whether other arms (like Perseus or Sagittarius) have different 
corotation radii, the cluster sample is specifically reacting to the radial gravitational 
gradient of the Local Arm.
The clusters with relative large misalignment are mostly young (log(age /yr) $<$ 8.25), 
with some exceptions. As can be seen in the Galactic
coordinate planes (similar to Figure~\ref{fig1} provided as online Supplementary material)
their tidal tails do not seem to  follow the typical S-shape of the nearer tidal tail region
of the clusters; they represent the total clusters' tidal tails length. 

While the misaligned population is predominantly composed of younger 
systems, Figure~\ref{fig5} reveals that the length of the tidal tail is the more 
direct physical indicator of alignment. Specifically, the transition at 
$\sim$0.5--0.6~kpc suggests that as tails extend beyond the immediate 
influence of the cluster’s Lagrange points, they transition from being 
tracers of the local radial potential to being tracers of global Galactic
 shear. This implies that tail length, rather than age alone, serves as 
the critical parameter for probing the zero-shear environment of the
 corotation resonance.

I propose that these open clusters could serve as instantaneous tracers of 
the local gravitational potential, where their relatively short tails ($< 0.5$ kpc) have 
not yet been subjected to the long-term, time-averaged smoothing of global Galactic shear. 
This interpretation is supported by the outside-in nature of stellar
 stripping in energy space; stars with the lowest binding energy are the first to escape 
through the Lagrange points, initially following trajectories dictated by the raw radial 
perturbations of the Local Arm potential \citep{choietal2007}. In the specific zero-shear
 zone of the corotation resonance, where azimuthal stretching is suppressed, these initial
 exit paths dominate the observed geometry \citep{barrosetal2021}. However, as clusters 
age beyond $\sim$ 200 Myr, they typically migrate $\sim$ 1-2 kpc from their birth 
positions \citep{babaetal2024}. This migration timescale coincides with the observed 
vanishing of large $\Delta$PA values, indicating that as clusters move out of the resonance 
trap or undergo multiple disc crossings, the cumulative effect of Galactic shear 
eventually stretches the tails into standard tangential arcs \citep{ramezanietal2026}. 
Consequently, the misaligned population represents a transient dynamical phase, catching 
these systems in a raw state before orbital phase-mixing erases the signature of the 
spiral arm resonance.

\begin{figure}
\includegraphics[width=\columnwidth]{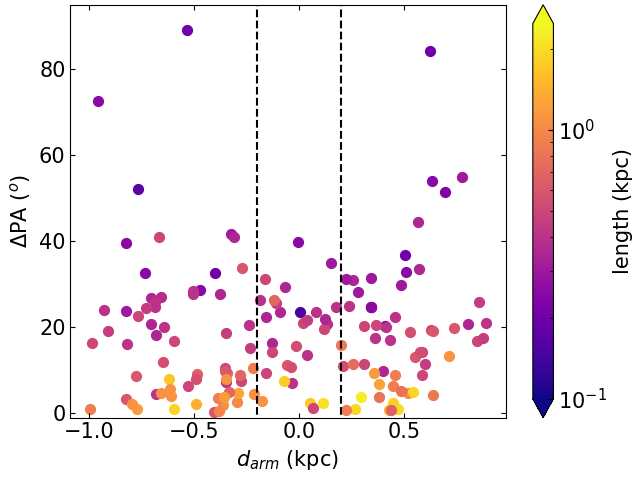}
\caption{ Relationship between $\Delta$PA and the 
distance of the open clusters to the Local Arm ($d_{arm}$). The points are 
coloured according to  the length of the open cluster tidal tails. The vertical dashed lines
represent the boundaries of the Local Arm \citep{haoetal2021}.}
\label{fig6}
\end{figure}

I superimpose in Figure~\ref{fig4} the mean position and physical width of the Local 
Arm as defined by \citet{haoetal2021}. To quantify the spatial distribution of these 
orientations, I computed the distance of each open cluster to the Local Arm ($d_{arm}$) 
and built Figure~\ref{fig6}, which illustrates the relationship between $\Delta$PA and 
$d_{arm}$. As can be seen, misalignments of $\Delta$PA $\gtrsim$ 20$\degr$ are observed
 across the entire surveyed extension of the solar neighborhood. However, the most 
significant departures from orbital alignment ($\Delta$PA $\gtrsim$ 40$^o$) occur 
symmetrically at both sides of the Local Arm boundaries, effectively placing these highly
 misaligned clusters in the inter-arm gaps. 

Figure~\ref{fig5} provides the physical context for these anomalies: the clusters
 exhibiting misalignments larger than 40$^o$ are characterized by the shortest tidal 
tails in the sample, with physical lengths $\lesssim$ 0.3~kpc. This confirms that these
 nascent structures act as high-precision, instantaneous tracers of the raw radial 
gravitational forcing within the zero-shear zone of the corotation resonance
 \citep{barrosetal2021}. In these inter-arm regions, the suppression of standard azimuthal
 stretching allows the Local Arm potential to dominate the exit trajectories of stars 
from the Lagrange points before significant length is achieved. As the tails extend 
beyond $\sim$0.5--0.6~kpc, they transition into a shear-dominated regime where they 
eventually align with the cluster's orbital motion, as seen in the more evolved systems in Figure~\ref{fig6} \citep{ramezanietal2026, babaetal2024}. Note that while the inner 
regions of long tails are indeed locally perturbed (leading to S-shaped structures), 
their global linear fits are dominated by the shear-aligned outer structures (see Sec. 2).

The gradient of the Milky Way potential would seem to
become more complex the further clusters get from the corotation equilibrium point.
At the corotation radius, the standard azimuthal shear, which typically aligns tails with 
the orbit, is suppressed, allowing the raw radial gradient of the Local Arm to dominate the tidal stripping.
When a cluster moves away from a stagnation point, the first significant non-zero derivative of
 the Taylor expansion of the potential is often the third derivative. This happens when the tidal tensor
is not constant, causing the tidal force scales with the square of the distance and, crucially, to point 
in a radial direction toward or away from the spiral arm.
 \citep[see, e.g.,][]{bw1967,barbanis1970,masset2000}. Being the Sun close to the
corotation radius \citep{barrosetal2021}, I would expected that open clusters further to it
had had their tidal tails not aligned with their direction of motion.

The open clusters with significantly non-aligned tidal tails ($\Delta$PA $\gtrsim$ 40$^o$)
 are distributed symmetrically at both sides of the Local Arm (see Figure~\ref{fig6}), 
effectively placing them in the inter-arm gaps. While a baseline misalignment of $\Delta$PA
 $\gtrsim$ 20$^o$ is detected across the entire surveyed range of $d_{arm}$, these 
extreme departures represent a localized dynamical signature that cannot be attributed to 
measurement noise. Given that our calculated PA uncertainties are only $\pm 5\degr$
(derived from the membership-threshold sensitivity tests in Section 2.1), 
these misalignments are nearly eight times larger than the uncertainty floor.
 The statistical significance of this result is further underscored by the 94.5\% accuracy of
the {\it Gaia} DR3 cluster census within 1~kpc, ensuring that these anomalies are not artifacts of 
selection effects \citep{huntetal2026}. Notably, clusters with aligned tidal tails in these 
same inter-arm regions are typically older, suggesting they have had sufficient time to 
conform their structures to the global orbital motion.

The placement of these highly misaligned clusters suggests that the phenomenon is driven by 
the unique dynamical conditions of the inter-arm zones at the corotation radius. In these 
specific locations, the suppression of standard azimuthal Galactic shear exposes clusters to 
the raw radial gravitational gradient of the Local Arm potential, which dominates the 
stripping geometry \citep{barrosetal2021}. This results in a transient phase restricted to 
the nascent stages of tail development; as Figure~\ref{fig5} confirms, the largest 
misalignments are found only in clusters with physical tail lengths $\lesssim 0.3$ kpc. As 
these systems evolve beyond $\sim$200 Myr and migrate $\sim$1-2 kpc out of the resonance trap,
 global shear eventually aligns the tails with the orbital motion 
\citep{babaetal2024, ramezanietal2026}. Conversely, the presence of relatively short but 
aligned tails in older
 open clusters likely reflects the cumulative effect of multiple disc crossings, which
 periodically reset tidal structures to align with the time-averaged global 
potential \citep{kupperetal2010, jerabkovaetal2021}.

The discovery of a population of clusters with significant tidal tail misalignments 
($\Delta$PA $\gtrsim$ 40$^o$) specifically in the inter-arm gaps of the Local Arm 
suggests a localized resonant driver rather than a purely global Galactic effect. 
While the baseline misalignment observed across the broader sample ($\Delta$PA $\approx$ 
20$^o$) may be attributed to large-scale disc features, the 
 nascent tails ($\lesssim$ 0.3 kpc) require a multi-perturber framework \cite{kosetal2026}.

A granular analysis of Figure~\ref{fig6} reveals that the relationship between tidal tail 
misalignment and the Local Arm would seem governed by three distinct dynamical regimes. These 
regimes could suggest that stellar stripping is influenced by a nested hierarchy of Galactic 
perturbers \citep{paruletal2026, kosetal2026}.
The first two regimes exhibit a visible spatial indifference to the Local Arm boundaries. 
Older clusters, characterized by evolved tidal tails, remain predominantly aligned ($\Delta$PA 
$\approx$ 0$^o$) across the entire surveyed range of $d_{arm}$. Similarly, younger clusters 
with moderate misalignments (10$^o \le \Delta$PA $\le$ 30$^o$) are distributed uniformly both 
within the Local Arm and across the inter-arm gaps. This behavior indicates that these 
populations are reacting to a global dynamical driver, the Galactic bar, whose gravitational 
torques induce systematic deflection angles that are robust over long-term orbital evolution 
and indifferent to local spiral features \citep{paruletal2026}.

The alignment of the benchmark clusters Melotte 25 (Hyades) and NGC 2632 (Praesepe) provides 
a critical constraint for this global regime. As shown in Table 1, Melotte 25 ($\Delta$PA 
$\approx$ 7$^o$) and NGC 2632 ($\Delta$PA $\approx$ 2.4$^o$) are well-aligned with their 
orbital motion. Despite being situated near the corotation radius, their alignment disfavours 
moderate Galactic bar pattern speeds ($\Omega_b \approx$ 39 - 45 km s$^{-1}$ kpc$^{-1}$), which 
would otherwise induce deflection angles inconsistent with the observed data 
\citet{paruletal2026}. Their alignment could indicate they reside in regions where the 
global disk shear is sufficient to balance local perturbations.

On the other hand, the third regime consists of clusters with large misalignments ($\Delta$PA 
$\gtrsim$ 40$^o$), which exhibit a unique spatial specificity. These clusters are found 
exclusively in the inter-arm gaps ($|d_{arm}| \gtrsim$ 0.2 kpc), appearing in similar numbers 
on both sides of the Local Arm. The fact that the 10$^o$ - 30$^o$ population behaves as if 
the Local Arm would not exist reinforces the speculation that the extreme $\Delta$PA
($>$40$^o$) would be a unique byproduct of the spiral potential. Within these gaps, the 
suppression of azimuthal shear creates a zero-shear zone where nascent tails are exposed to 
the raw radial gravitational forcing of the Local Arm resonance before they grow long 
enough to be captured by the steady-state global disk flow \citep{barrosetal2021, kosetal2026}. 

The high susceptibility of nascent tidal tails ($\le$ 0.3 kpc) to the Local Arm resonance can 
be explained by the transition from ballistic escape to a shear-dominated regime. 
The escape of stars from an open cluster is dictated by the tidal tensor, $\mathbf{T}_{ij}$, 
evaluated at the Lagrangian points $L_1$ and $L_2$ \citep{kos2024}. Within the zero-shear zone 
of the Local Arm corotation resonance, the suppression of azimuthal stretching allows 
third-order derivatives of the spiral potential to become the dominant terms in the local 
expansion \citep{barrosetal2021}. This leads to a physical rotation of the tidal tensor's 
eigenvectors toward a radial orientation. Stars in nascent tails are still aligned with 
these rotated local eigenvectors, whereas stars in evolved tails have drifted far enough 
from the core to integrate the larger-scale disk potential, which is dominated by the 
steady-state rotation curve \citep{paruletal2026}.

I propose that the return to alignment seen in Figure~\ref{fig5} could represent a shear 
relaxation timescale. In this framework, Galactic differential rotation would act as a 
restorative torque that gradually realigns the stellar spine with the orbital velocity 
vector \citep{kosetal2026}. For the inter-arm gaps, Table~\ref{tab1}'s data identifies 
a critical relaxation length of $L_{rel} \approx$ 0.4 - 0.5 kpc. Below this threshold, 
the tail is a high-resolution probe of the instantaneous non-axisymmetric field; above 
it, the structure integrates the disk's gravitational potential over a significant orbital 
arc, effectively averaging out localized resonant perturbations \citep{paruletal2026}. 
This could identify nascent tails as uniquely sensitive sensors for mapping high-frequency 
fluctuations in the Galactic potential that are otherwise masked by the shear-averaged 
geometry of older systems.

\section{Conclusions}

In this work, I characterized the structural properties of tidal tails for a sample of 
155 nearby open clusters to investigate the influence of the Milky Way potential on stellar 
stripping. The main findings are summarized as follows:\\

$\bullet$ By employing a membership threshold of $P >$ 50$\%$, I traced physically reliable 
tidal tails, finding that their lengths generally correlate with open cluster age, with 
physical extension serving as the primary indicator of dynamical alignment.\\
    
$\bullet$ I identified three distinct dynamical regimes of tail geometry. The majority of 
the population follows the global Galactic shear, while a significant subset exhibits a 
spatially indifferent background misalignment (10$^o$ - 30$^o$) driven 
by the global torques of the Galactic bar.\\
    
$\bullet$ The alignment of benchmark clusters Melotte 25 and NGC 2632 ($\Delta$PA $<$ 
10$^o$) provides a critical empirical constraint, disfavouring moderate Galactic bar 
pattern speeds ($\Omega_b \approx$ 39 - 45 km s$^{-1}$ kpc$^{-1}$) which would 
otherwise induce significant deflections in these regions.\\
    
$\bullet$ I discovered a localized dynamical signature of open clusters whose nascent tidal 
tails ($\le$ 0.3 kpc) are highly misaligned ($\Delta$PA $\gtrsim$ 40$^o$). This 
resonant signal is situated symmetrically in the inter-arm gaps ($|d_{arm}| \gtrsim$ 0.2 kpc)
 beyond the Local Arm boundaries.\\
    
$\bullet$ These anomalies occur within the corotation resonance's zero-shear zone, 
where standard azimuthal disk shear is suppressed. In this environment, the radial 
gravitational forcing of the spiral potential dominates the stellar exit trajectories by 
rotating the eigenvectors of the tidal tensor.\\
    
$\bullet$ This misalignment represents a transient dynamical phase. Nascent tails could serve 
as high-resolution, instantaneous tracers of the non-axisymmetric potential, while 
evolved tails ($\gtrsim$ 0.6 kpc) integrate the disk's potential over larger orbital arcs,
 eventually relaxing into alignment with the steady-state Galactic rotation curve.

\begin{acknowledgements}
I thank the referee for the thorough reading of the manuscript and
timely suggestions to improve it.

Data used in this work are available upon request to the author.

\end{acknowledgements}


\begin{table*}
\caption{PA and length of tidal tails of selected open clusters.} 
\label{tab1}
\begin{tabular}{lcccclcccc}
\hline
Name & log($t$ /yr) & PA$_{tail}$ & length & PA$_{vel}$ & Name & log($t$ /yr) & PA$_{tail}$ & length & PA$_{vel}$\\
          &              & (deg) & (kpc) &  (deg)   &       &             & (deg) & (kpc) &  (deg)  \\   
\hline
UPK~350 & 8.10 & 65.00 & 0.25 & 89.72 & NGC~6087 & 8.00 & 160.00 & 0.25 & 87.36 \\
UPK~55 & 8.32 & 54.00 & 0.25 & 82.74 & Collinder~338 & 8.15 & 60.00 & 0.25 & 92.49 \\
UPK~508 & 8.00 & 60.00 & 0.25 & 99.78 & UPK~467 & 8.06 & 60.00 & 0.30 & 91.12 \\
NGC~2632 & 8.82 & 100.00 & 1.72 & 97.63 & Mamajek~4 & 8.56 & 70.00 & 0.40 & 97.80 \\
UPK~433 & 8.25 & 60.00 & 0.30 & 89.93 & UPK~51 & 8.02 & 65.00 & 0.35 & 92.63 \\
Alessi~1 & 9.15 & 70.00 & 0.20 & 90.32 & NGC~1758 & 8.43 & 65.00 & 0.35 & 85.69 \\
UPK~305 & 8.14 & 60.00 & 0.30 & 88.08 & UPK~167 & 8.08 & 65.00 & 0.35 & 86.87 \\
UBC~255 & 9.37 & 85.00 & 0.50 & 92.60 & UPK~53 & 8.65 & 85.00 & 0.90 & 88.67 \\
Gulliver~20 & 8.24 & 80.00 & 0.40 & 87.37 & UPK~612 & 8.00 & 55.00 & 0.60 & 88.66 \\
Gulliver~11 & 8.50 & 70.00 & 0.50 & 86.70 & NGC~2423 & 9.03 & 85.00 & 1.80 & 87.38 \\
UPK~644 & 8.22 & 70.00 & 0.30 & 93.88 & Melotte~111 & 8.80 & 80.00 & 1.80 & 87.50 \\
ASCC~101 & 8.68 & 80.00 & 0.50 & 89.30 & NGC~2287 & 8.22 & 70.00 & 0.50 & 90.36 \\
Stock~23 & 8.03 & 60.00 & 0.40 & 82.35 & ASCC~23 & 8.36 & 65.00 & 0.55 & 83.93 \\
UPK~524 & 8.02 & 70.00 & 0.35 & 93.54 & Alessi~Teutsch~11 & 8.15 & 65.00 & 0.50 & 86.00 \\
NGC~2168 & 8.16 & 30.00 & 0.30 & 84.79 & NGC~1647 & 8.56 & 80.00 & 0.80 & 85.14 \\
ASCC~10 & 8.41 & 70.00 & 0.60 & 83.08 & NGC~1662 & 8.88 & 80.00 & 2.30 & 83.77 \\
Collinder~350 & 8.76 & 88.00 & 1.10 & 89.90 & NGC~1750 & 8.40 & 65.00 & 0.60 & 84.21 \\
Ruprecht~161 & 8.01 & 65.00 & 0.20 & 97.55 & UPK~429 & 8.14 & 45.00 & 0.35 & 89.34 \\
ASCC~99 & 8.54 & 85.00 & 0.70 & 94.39 & Platais~3 & 8.24 & 90.00 & 0.50 & 91.20 \\
NGC~2925 & 8.10 & 75.00 & 0.40 & 95.46 & UPK~180 & 8.38 & 75.00 & 0.50 & 86.39 \\
NGC~2682 & 9.62 & 100.00 & 1.97 & 99.08 & NGC~2422 & 8.03 & 70.00 & 0.40 & 94.60 \\
Alessi~Teutsch~3 & 8.01 & 70.00 & 0.45 & 95.62 & IC~4725 & 8.04 & 65.00 & 0.35 & 83.23 \\
Stock~5 & 8.11 & 135.00 & 0.23 & 83.61 & UPK~4 & 8.43 & 65.00 & 0.40 & 81.03 \\
UPK~431 & 8.77 & 85.00 & 0.90 & 91.27 & UPK~84 & 9.00 & 70.00 & 1.25 & 74.69 \\
UPK~560 & 8.50 & 80.00 & 0.45 & 95.24 & UPK~537 & 8.70 & 80.00 & 0.50 & 94.27 \\
UPK~645 & 8.01 & 36.00 & 0.17 & 88.05 & UPK~45 & 8.67 & 45.00 & 0.50 & 86.06 \\
UPK~333 & 8.87 & 90.00 & 0.85 & 94.15 & NGC~6025 & 8.81 & 65.00 & 0.50 & 87.55 \\
UPK~578 & 8.93 & 95.00 & 0.40 & 99.56 & NGC~1708 & 8.32 & 60.00 & 0.45 & 85.86 \\
NGC~6281 & 8.70 & 95.00 & 1.10 & 91.05 & NGC~2548 & 8.77 & 80.00 & 1.20 & 89.31 \\
NGC~7092 & 8.59 & 65.00 & 0.55 & 80.58 & Pismis~4 & 8.80 & 65.00 & 0.35 & 94.45 \\
Alessi~9 & 8.44 & 80.00 & 0.70 & 88.98 & COIN-Gaia~1 & 8.52 & 65.00 & 0.40 & 82.44 \\
Alessi~6 & 8.62 & 90.00 & 0.60 & 98.74 & UPK~12 & 8.41 & 65.00 & 0.40 & 85.04 \\
Stock~1 & 8.61 & 85.00 & 1.05 & 89.38 & COIN-Gaia~12 & 8.59 & 65.00 & 0.40 & 85.94 \\
Gulliver~21 & 8.43 & 85.00 & 0.55 & 95.92 & Ferrero~1 & 8.62 & 65.00 & 0.40 & 88.96 \\
NGC~6475 & 8.34 & 80.00 & 0.50 & 90.61 & Alessi~37 & 8.68 & 70.00 & 0.85 & 85.77 \\
UBC~8 & 8.69 & 80.00 & 1.25 & 86.70 & ASCC~88 & 8.53 & 75.00 & 0.50 & 91.28 \\
NGC~6124 & 8.27 & 65.00 & 0.35 & 91.75 & UPK~5 & 8.42 & 70.00 & 0.50 & 86.78 \\
NGC~5460 & 8.19 & 65.00 & 0.37 & 91.92 & NGC~6494 & 9.05 & 90.00 & 1.20 & 88.89 \\
UPK~143 & 8.40 & 65.00 & 0.40 & 85.72 & UPK~54 & 8.52 & 55.00 & 0.40 & 83.35 \\
Trumpler~2 & 8.04 & 55.00 & 0.25 & 87.91 & COIN-Gaia~13 & 8.67 & 85.00 & 0.65 & 84.30 \\
Lynga~2 & 8.00 & 135.00 & 0.25 & 95.47 & Loden~1194 & 8.38 & 85.00 & 0.55 & 96.85 \\
Alessi~2 & 8.54 & 80.00 & 1.00 & 84.80 & Alessi~44 & 8.71 & 80.00 & 0.55 & 88.05 \\
NGC~752 & 9.06 & 92.00 & 1.90 & 90.97 & Ferrero~11 & 8.33 & 75.00 & 0.55 & 89.32 \\
IC~4651 & 9.21 & 95.00 & 0.90 & 96.00 & ASCC~113 & 8.45 & 75.00 & 0.55 & 85.71 \\
UBC~32 & 8.89 & 50.00 & 0.30 & 91.55 & ASCC~128 & 8.39 & 75.00 & 0.75 & 86.44 \\
Alessi~31 & 8.37 & 60.00 & 0.40 & 84.64 & NGC~2358 & 8.81 & 75.00 & 0.40 & 92.11 \\
IC~4756 & 9.10 & 90.00 & 0.60 & 90.36 & ASCC~90 & 8.90 & 90.00 & 1.10 & 85.23 \\
NGC~6997 & 8.80 & 65.00 & 0.40 & 78.65 & UPK~567 & 8.52 & 90.00 & 0.95 & 90.51 \\
NGC~2184 & 8.80 & 92.00 & 0.80 & 95.77 & NGC~6991 & 9.18 & 70.00 & 0.15 & 93.64 \\
NGC~7243 & 8.16 & 60.00 & 0.40 & 85.04 & NGC~1901 & 8.94 & 90.00 & 1.10 & 92.93 \\
NGC~3532 & 8.59 & 90.00 & 1.00 & 92.54 & UPK~27 & 9.20 & 90.00 & 0.60 & 93.26 \\
Stock~12 & 8.04 & 55.00 & 0.30 & 90.00 & Blanco~1 & 8.01 & 65.00 & 0.70 & 91.37 \\
NGC~1545 & 8.03 & 0.00 & 0.20 & 84.14 & NGC~2527 & 8.83 & 100.00 & 2.10 & 97.51 \\
NGC~2281 & 8.78 & 92.00 & 1.40 & 92.84 & ASCC~11 & 8.38 & 65.00 & 0.60 & 84.73 \\
UPK~194 & 8.17 & 75.00 & 0.35 & 84.77 & NGC~1342 & 8.90 & 80.00 & 1.90 & 84.85 \\
ASCC~111 & 8.43 & 70.00 & 0.35 & 86.29 & UPK~136 & 8.05 & 65.00 & 0.45 & 86.68 \\
NGC~5662 & 8.29 & 75.00 & 0.35 & 95.77 & NGC~6633 & 8.83 & 100.00 & 1.10 & 92.13 \\
UPK~237 & 8.17 & 65.00 & 0.40 & 85.09 & COIN-Gaia~30 & 8.40 & 70.00 & 0.90 & 78.90 \\
UPK~579 & 8.07 & 5.00 & 0.20 & 94.09 & UPK~442 & 8.22 & 60.00 & 0.30 & 91.35 \\
Alessi~12 & 8.12 & 60.00 & 0.40 & 86.41 & Ruprecht~147 & 9.47 & 70.00 & 0.70 & 74.90 \\
RSG~1 & 8.08 & 60.00 & 0.35 & 91.08 & Stock~2 & 8.59 & 95.00 & 0.90 & 94.33 \\
ASCC~73 & 8.52 & 87.00 & 0.65 & 96.16 & NGC~2301 & 8.32 & 80.00 & 0.45 & 91.49 \\
NGC~2516 & 8.37 & 70.00 & 0.35 & 92.41 & COIN-Gaia~23 & 8.10 & 70.00 & 0.45 & 87.46 \\
COIN-Gaia~11 & 8.89 & 80.00 & 0.50 & 88.89 & NGC~2546 & 8.14 & 70.00 & 0.40 & 93.54 \\
\hline
\end{tabular}
\end{table*}

\setcounter{table}{0}
\begin{table*}
\caption{continued.} 
\label{tab1}
\begin{tabular}{lcccclcccc}
\hline
Name & log($t$ /yr) & PA$_{tail}$ & length & PA$_{vel}$ & Name & log($t$ /yr) & PA$_{tail}$ & length & PA$_{vel}$\\
          &              & (deg) & (kpc) &  (deg)   &       &             & (deg) & (kpc) &  (deg)  \\   
\hline
Alessi~62 & 8.83 & 75.00 & 0.60 & 84.83 & Gulliver~28 & 8.52 & 65.00 & 0.45 & 83.64 \\
Alessi~8 & 8.06 & 70.00 & 0.40 & 95.78 & Harvard~10 & 8.29 & 70.00 & 0.45 & 94.44 \\
UPK~381 & 8.45 & 70.00 & 0.50 & 89.33 & NGC~5822 & 8.95 & 100.00 & 1.10 & 97.80 \\
King~6 & 8.30 & 70.00 & 0.50 & 84.36 & Ruprecht~98 & 8.68 & 85.00 & 1.20 & 88.75 \\
NGC~2215 & 8.83 & 30.00 & 0.25 & 83.88 & UPK~29 & 8.90 & 75.00 & 1.70 & 83.06 \\
ASCC~41 & 8.32 & 65.00 & 0.50 & 84.54 & NGC~1039 & 8.11 & 65.00 & 0.50 & 85.42 \\
UPK~20 & 9.32 & 87.00 & 1.40 & 89.14 & Teutsch~35 & 8.00 & 55.00 & 0.50 & 86.19 \\
COIN-Gaia~25 & 8.74 & 70.00 & 1.10 & 83.27 & UPK~21 & 9.05 & 95.00 & 1.90 & 93.92 \\
ESO~130~06 & 8.14 & 90.00 & 0.50 & 96.28 & UPK~93 & 8.70 & 70.00 & 0.90 & 80.59 \\
UPK~220 & 8.03 & 120.00 & 0.20 & 83.27 & Ruprecht~145 & 9.05 & 95.00 & 1.10 & 100.69 \\
Collinder~463 & 8.05 & 50.00 & 0.35 & 83.54 & UPK~13 & 8.52 & 65.00 & 0.45 & 84.08 \\
NGC~6793 & 8.48 & 50.00 & 0.35 & 91.07 & UPK~296 & 8.31 & 70.00 & 0.60 & 88.03 \\
Alessi~3 & 8.79 & 80.00 & 0.60 & 91.26 & Melotte~22 & 7.88 & 50.00 & 0.20 & 88.46 \\
Melotte~25 & 8.89 & 90.00 & 0.40 & 96.99 &  & & & &  \\
\hline
\end{tabular}
\end{table*}

\end{document}